# $\delta^{44/40}$Ca-$\delta^{88/86}$Sr multi-proxy constrains primary origin of Marinoan cap carbonates

**Jiuyuan Wang[1,2], Andrew D. Jacobson[1], Bradley B. Sageman[1], Matthew T. Hurtgen[1]**

*1 Department of Earth and Planetary Sciences, Northwestern University, Evanston, IL, USA, 60208*

*2 Department of Earth and Planetary Sciences, Yale University, New Haven, CT, USA, 06520*

*Email address: jiuyuan.wang@yale.edu*

## Abstract

The Neoproterozoic Earth experienced at least two global-scale glaciations termed Snowball Earth events. "Cap carbonates" were widely deposited after the events, but controversy surrounds their origin. Here, we apply the novel $\delta^{44/40}$Ca-$\delta^{88/86}$Sr multi-proxy to two Marinoan (ca. 635 Ma) cap carbonate sequences from Namibia and show that the rocks archive primary environmental signals deriving from a combination of seawater-glacial meltwater mixing and kinetic isotope effects. In an outer platform section, dolostone $\delta^{44/40}$Ca and $\delta^{88/86}$Sr values define a line predicted for kinetic mass-dependent isotope fractionation. This dolostone mostly precipitated from meltwater. Moreover, stratigraphically higher samples exhibiting the fastest precipitation rates correlate with elevated $^{87}$Sr/$^{86}$Sr ratios, consistent with long-held expectations that a rapid deglacial weathering pulse forced cap carbonate formation. An inner-platform dolostone shows greater effects from water-mass mixing but still reveals that precipitation rates increased up-section. Overlying limestones show the greatest Ca and Sr contributions from seawater. Amplification of local coastal processes during global ice sheet collapse offers a simple but sufficient proposition to explain the Ca isotope heterogeneity of cap carbonates. Detection of kinetic isotope effects in the rock record provides a basis for developing the $\delta^{44/40}$Ca-$\delta^{88/86}$Sr multi-proxy as an indicator of saturation state and $p$CO$_2$.

**Introduction**

"Snowball Earth" events during the Cryogenian period (ca. 635 – 720 Ma) represent some of the planet's most extreme forms of climate change. Ice sheet collapse following the events likely led to major environmental upheaval, as recorded by the widespread deposition of "cap carbonate" sequences[1] bearing large carbon isotope ($\delta^{13}$C) variations[2-4] and sedimentary evidence for rapid marine transgression[5, 6]. The globally-distributed cap carbonates, which directly overlie glacial deposits, comprise transgressive "cap dolostones"[6-8] followed by limestones in many locations[9]. Multiple mechanisms have been proposed to explain the origin of cap carbonates and the geochemical signals contained within them, including water-mass mixing[10, 11], gas hydrate destabilization[12], ocean warming[13], sediment starvation[14], loess weathering[15], and diagenetic transformation[16], although interpretations remain controversial.

The calcium isotope proxy ($\delta^{44/40}$Ca) offers a valuable tool for studying cap carbonates, but existing applications have yielded divergent interpretations representing nearly the full range of possibilities. Cap carbonate $\delta^{44/40}$Ca records have been attributed to global-scale imbalances between riverine runoff and carbonate burial[17], local-scale increases in continental chemical weathering rates[18], and spatial and temporal shifts between rock- versus fluid-buffered early diagenesis[16]. Noteworthy is that kinetic isotope effects appear to explain $\delta^{44/40}$Ca variability in a variety of Phanerozoic records[19-22], including those generated from shallow-water carbonates[22], but this mechanism has not yet been demonstrated in Neoproterozoic studies. Deciphering the Ca isotope geochemistry of cap carbonates has numerous implications, including interpreting Neoproterozoic $\delta^{13}$C excursions[3, 4], identifying whether cap dolostones are primary or secondary precipitates[7, 23, 24], and otherwise reconstructing environmental conditions during major climate catastrophes[7, 25].

Here, we show for the first time that the complementary analysis of stable Sr isotope ratios ($\delta^{88/86}$Sr) can constrain the fidelity of cap carbonate $\delta^{44/40}$Ca records and further elucidate their meaning. The "$\delta^{44/40}$Ca-$\delta^{88/86}$Sr multi-proxy" traces both mass-dependent fractionation and reservoir mixing, analogously to other three- and four-isotope systems used to study problems in terrestrial biogeochemistry and cosmochemistry[26]. While these other isotope systems pertain to single elements, the four-isotope approach presented here capitalizes on well-known observations that Ca and Sr participate in many of the same geochemical reactions and exhibit similar behavior. In detail, the proxy may be considered a five-isotope system, as quantification of $\delta^{88/86}$Sr values includes analysis of fractionation-corrected radiogenic Sr isotope ratios ($^{87}$Sr/$^{86}$Sr), which provide additional constraints on mixing.

If variable mass-dependent fractionation was the only source of Ca and Sr isotopic variability during carbonate formation, then samples will define a mass-fractionation line exhibiting a characteristic slope. Mass-fractionation laws predict that carbonate $\delta^{88/86}$Sr and $\delta^{44/40}$Ca values should linearly covary with a slope of 0.12 for equilibrium control, or 0.19 and 0.24 for kinetic control[27]. If mixing (e.g., global-scale flux perturbations, local-scale freshwater inputs, or some combination thereof) drove changes in the isotopic composition of seawater at timescales faster than the rate of carbonate accumulation, then carbonates will not yield a mass-fractionation line. Further, if carbonates were subject to early diagenetic alteration, whether recrystallization of primary calcite or secondary addition of authigenic calcite, then carbonates will also deviate from an ideal mass-fractionation line. Here, however, samples can only trend toward higher $\delta^{44/40}$Ca at nearly constant $\delta^{88/86}$Sr, mainly because primary carbonates have much higher

Sr/Ca ratios than secondary carbonate[16, 28], and as with Ca isotopes[29], no Sr isotope fractionation accompanies diagenetic calcite formation[27]. In addition, observations and models focused on modern and ancient sedimentary systems show that the $\delta^{88/86}$Sr proxy is largely insensitive to alteration because primary carbonates and fluids have limited isotopic contrast and any re-equilibration between them is exceedingly small[22, 30].

In this contribution, we apply the $\delta^{44/40}$Ca-$\delta^{88/86}$Sr multi-proxy to cap carbonates deposited after the Marinoan glaciation (ca. 635 Ma). Stable Ca and Sr isotope records independently exist for Marinoan cap carbonates, but their interpretations fundamentally differ. An early investigation reporting $\delta^{88/86}$Sr values interpreted signals as primary[31], whereas a more recent study reporting $\delta^{44/40}$Ca values interpreted signals as representing varying degrees of rock- vs. fluid-buffered diagenetic alteration[16]. The multi-proxy results presented herein resolve this conundrum by demonstrating that the combined signals are high fidelity and derive from a combination of seawater-glacial meltwater mixing and kinetic isotope effects. These mechanisms readily explain the extreme heterogeneity commonly observed for cap carbonate $\delta^{44/40}$Ca records and could have contributed to $\delta^{13}$C variations as well.

**Results**

We investigated cap carbonate sequences from the Ombaatjie (P4017, n=16) and Arbeitsgenot (G2008, n=21) sections in Namibia (Figure 1). The Maieberg Formation overlies the diamictitic Ghaub Formation and comprises a basal member of dolostone with columnar stromatolitic structures and wave ripples (Keilberg cap dolostone), a middle member of limestone rhythmite (here, termed the cap limestone), and an upper member of dolomite[9, 23]. The sections were likely deposited on a rimmed carbonate platform situated in subtropical latitudes[32], with the Arbeitsgenot section located much closer to the outer edge of the platform than the Ombaatjie section[6, 9, 33]. Section correlation was adopted from previous studies using $\delta^{13}C$ chemostratigraphy[33, 34]. All Ca and Sr isotope ratios ($\delta^{44/40}Ca$, $\delta^{88/86}Sr$, $^{87}Sr/^{86}Sr$), as well as other elemental concentrations, were measured at Northwestern University. Resulting $\delta^{44/40}Ca$ and $\delta^{88/86}Sr$ values are reported relative to the OSIL Atlantic seawater standard and the NIST-987 strontium carbonate standard, respectively. Select samples were also measured for $\delta^{44/42}Ca$ values, to assess potential radiogenic $^{40}Ca$ enrichments (see Supplementary information for details).

Measured $\delta^{44/40}Ca$ range from -0.75‰ to -1.36‰ (Figure 2). Both sections display negative trends within the Keilberg cap dolostones followed by positive shifts in the cap limestones immediately above the flooding surface, a pattern consistent with previous observations[16, 18, 35]. $\delta^{88/86}Sr$ range from 0.19‰ to 0.51‰, with the cap dolostones displaying less variation than the cap limestones. The upper part of the cap limestone in the Arbeitsgenot section has $\delta^{88/86}Sr$ values as high as 0.51‰, which matches the end-Ediacaran background[36]. $^{87}Sr/^{86}Sr$ ratios steadily increase from ~0.7085 at the base of the Keilberg cap dolostones to values as high as 0.7119 in the overlying limestones. Other studies of Marinoan cap carbonates have reported background $^{87}Sr/^{86}Sr$ ratios near 0.7085, with some dolostone ratios approaching 0.7123[35]. Radiogenic $^{40}Ca$ enrichments were not detected (Figure S1, see Supplementary information for detailed evaluation).

**Discussion**

*Primary versus secondary signals*

Similar to findings reported in previous cap carbonate studies[33], the present samples show ambiguous evidence for late-stage diagenetic effects. Although $\delta^{13}C$ and $\delta^{18}O$ generally covary, other proxy data show mixed signals (Figure S6-7). The upper portions of the Keilberg cap dolostones and lower portions of the cap limestones display elevated concentrations of Fe and Mn, as well as high $^{87}Sr/^{86}Sr$ ratios relative to those for globally well-mixed seawater[37]. High $^{87}Sr/^{86}Sr$ ratios are typically eliminated from consideration based on the assumption that they represent contamination from radiogenic fluids or detrital sediments[38], yet coastal carbonates commonly record $^{87}Sr/^{86}Sr$ ratios both higher and lower than global seawater due to local inputs from rivers and groundwater[39-41].

Early diagenetic effects are more probable. Recent studies have proposed that Ca isotopes mostly trace recrystallization and aragonite neomorphism under rock- versus fluid-buffered conditions[16, 42]. Rock-buffered alteration preserves primary $\delta^{44/40}Ca$ values, while fluid-buffered alteration elevates $\delta^{44/40}Ca$ values because no Ca isotope fractionation occurs when recrystallizing sediments equilibrate with seawater[29, 42]. Stable Sr isotope ratios, however, point to different controls. The cap limestones show an 0.33‰ spread in $\delta^{88/86}Sr$ values, but their relatively low and constant $\delta^{44/40}Ca$ values, as well as high Sr concentrations and low Mg concentrations, indicate these samples experienced the least alteration. Higher $\delta^{44/40}Ca$ values in the cap dolostones appear consistent with fluid-buffered aragonite neomorphism[16], yet $\delta^{88/86}Sr$ and $\delta^{44/40}Ca$ values for the Arbeitsgenot cap dolostone linearly covary with a slope of 0.210 (± 0.045, $R^2 = 0.74$, $p$-value = $1.5\times10^{-3}$), which is very close to the theoretical values of 0.19 and 0.24 predicted for chemical and diffusional kinetic isotope fractionations of aqueous $Ca^{2+}$ and $Sr^{2+}$. The "kinetic slope" has been recovered for diverse materials, including laboratory-synthesized inorganic calcites[27] (0.185±0.019, $R^2 = 0.96$, $p$-value = $6.0\times10^{-4}$), biologically-produced Cretaceous shallow water carbonates[22] (0.194±0.027, $R^2 = 0.66$, $p$-value = $7.7\times10^{-8}$), and waters and aragonite from the Coorong lagoon-estuarine system in South Australia[41] (0.162 ± 0.015, $R^2 = 0.93$, $p$-value = $5.2\times10^{-6}$), which greatly minimizes the possibility that the present result is a diagenetic artifact.

No evidence suggests that any form of diagenesis can increase $\delta^{88/66}Sr$ values while preserving primary $\delta^{44/40}Ca$ values. Similarly, mixing between primary and secondary carbonate cannot simulate the "kinetic slope," which is a mass-fractionation line (see Supplementary Information). As with Ca isotopes, little Sr isotope fractionation occurs when diagenetic calcite slowly forms under conditions of chemical equilibrium[27]. However, because slow calcite formation also maintains low Sr partition coefficients that inhibit Sr uptake and lead to Sr loss from recrystallizing sediments[27, 43, 44], the Sr/Ca ratios of diagenetic calcite are ~100 times lower than those for primary carbonate[28, 42]. The net result is that early diagenesis can affect $\delta^{44/40}$Ca but not $\delta^{88/86}$Sr[22, 30]. If anything, diagenesis should increase $\delta^{44/40}$Ca at nearly constant $\delta^{88/66}$Sr, but no such patterns are observed. We note that the $\delta^{88/66}$Sr and $\delta^{44/40}$Ca values for the Ombaatjie cap dolostone linearly covary with a slope of -0.106 ($\pm$ 0.048, $R^2$ = 0.41, $p$-value = 6.4×10$^{-2}$), which fits neither a simple kinetic model nor a diagenetic one. As demonstrated below, this pattern exemplifies the remaining prediction from the $\delta^{44/40}$Ca-$\delta^{88/86}$Sr multi-proxy, namely deviation due to water-mass mixing.

*A seawater-meltwater mixing origin of Marinoan cap carbonates*

An alternative hypothesis is needed to explain the dataset. The Marinoan deglaciation delivered a high volume of meltwater runoff to coastal regions[5, 8, 11], which suggests that the measured records likely comprise both global and local signals[18, 31, 45]. When meltwater mixes with seawater, the $\delta^{44/40}$Ca and $\delta^{88/86}$Sr values of the resulting mixtures are a function of the proportions of each water mass contributing to the mixtures, as well as their respective Ca and Sr concentrations and isotope ratios. These variables are difficult to precisely constrain for the Neoproterozoic. For example, seawater could have been brackish and possibly even a brine, and the exact composition of terrestrial runoff can only be inferred from the study of modern systems. Figure 3 presents an array of potential two-component mixing curves for the stable Ca and Sr isotope compositions of seawater-meltwater mixtures (see Supplementary Information), although determining the exact curve has little consequence for our interpretation.

In the four-isotope space mapped by the $\delta^{44/40}$Ca-$\delta^{88/86}$Sr multi-proxy, mass-fractionation during primary carbonate precipitation should produce samples that plot down and to the left from the theoretical mixtures (Figure 3). The slope measured for the Arbeitsgenot cap dolostone reveals kinetic fractionation of Ca and Sr isotopes, mostly from glacial meltwater. The relationships illustrated in Figure 3 show that the Arbeitsgenot dolostone began forming relatively near isotopic equilibrium with the meltwater-dominated mixture and that precipitation rates increased up-section. Interestingly, stratigraphically higher samples exhibiting the fastest precipitation rates also display higher $^{87}$Sr/$^{86}$Sr ratios. Post-glacial weathering mechanisms can substantially elevate the abundance of $^{87}$Sr in terrestrial runoff[46] independently of all other levers that control carbonate $\delta^{44/40}$Ca and $\delta^{88/86}$Sr values. Thus, we interpret the five-isotope system as providing support for long-held assertions that weathering-derived alkalinity from continental runoff forced rapid deposition of cap dolostones following ice-sheet collapse[3, 8, 10, 18, 47].

Recovery of the "kinetic slope" in the Arbeitsgenot cap dolostone implies that the net rate at which this unit formed exceeded the rate of mixing between meltwater and seawater, whereas absence of this result for this Ombaatjie cap dolostone suggests the opposite. This contrast presumably reflects differences in cross-platform position. The Ombaatjie cap dolostone formed in an inner platform environment, where the effects from seawater-meltwater mixing may have been more pronounced and possibly more sensitive to sea level rise. Regardless, comparing the best-fit slope defined by the Ombaatjie cap dolostone samples to the theoretical mixing curves between seawater and meltwater still reveals that precipitation rates increased up-section. Here too, faster precipitation rates broadly correlate with higher $^{87}$Sr/$^{86}$Sr ratios, which further points to a causal connection between enhanced chemical weathering and dolostone formation. Rather than indicating late-stage diagenetic alteration, we find that high $^{87}$Sr/$^{86}$Sr ratios commonly observed in cap carbonates offer primary environmental information[31, 35], which reinforces conclusions deduced from the behavior of stable Ca and Sr isotopes.

Figure 3 shows that the overlying cap limestones mostly precipitated from seawater, which is also consistent with sea level rise, as well as dissipation of the meltwater lens (See Supplementary Information). Radiogenic and stable Sr isotope ratios broadly negatively correlate, consistent with mixing between seawater and meltwater (see Figure S5). Elevated $^{87}$Sr/$^{86}$Sr ratios in some of the limestones only underscore that hydrologic discharge and chemical weathering during the glacial recession were highly dynamic and likely far from steady-state. While the behavior of the two stable isotope proxies suggests that the hydrological balance ultimately shifted more towards seawater as the cap carbonates formed, continental chemical weathering continued to elevate the abundance of $^{87}$Sr in runoff. Such is the case for limestone sample A-20, which has an $^{87}$Sr/$^{86}$Sr ratio of 0.7111. We note that another limestone sample (A-13) from the Arbeitsgenot section also has a high $^{87}$Sr/$^{86}$Sr ratio (0.7119) but lies outside the theoretically expected domain for carbonate mineral formation. No evidence suggests this sample manifests signals of alteration (see Figure S3). Instead, applying the "kinetic slope" to this datapoint suggests the sample precipitated from a seawater-meltwater solution that was not well-mixed and incompletely homogenized, i.e., the Ca isotope composition was more akin to seawater, while the Sr isotope composition (both radiogenic and stable) was more akin to meltwater. In either case, this pattern is clearly the exception, not the norm.

Lastly, it is noteworthy that the average $\delta^{88/86}$Sr value of all Maieberg carbonate samples (0.35‰) is substantially higher than that for Phanerozoic carbonates (0.16‰), which implies that the $\delta^{88/86}$Sr value of Neoproterozoic seawater was higher than modern seawater (0.39‰).

Carbonates from the end-Neoproterzoic[31, 36] and other time periods throughout Earth history[20] indicate that ancient seawater periodically achieved $\delta^{88/86}$Sr values higher than modern. Elevated $\delta^{88/86}$Sr values characterizing the Neoproterozoic ocean are consistent with a background steady-state where higher atmospheric $p$CO$_2$ sustained more rapid silicate weathering rates[37, 48] and thus, faster carbonate burial rates[3] compared to times with lower $p$CO$_2$. Carbonates spanning the Permian-Triassic boundary, also an interval of elevated $p$CO$_2$, point to high seawater $\delta^{88/86}$Sr values[49], which provides some support for the hypothesis that the late Permian and late Neoproterozoic oceans functioned similarly[12].

*Implications for the Marinoan $\delta^{13}$C record and end-Neoproterozoic carbon cycle*

We compiled all published $\delta^{44/40}$Ca records for cap dolostones from the Otavi platform and the Namibian foreslope and correlated them using $\delta^{13}$C chemostratigraphy[16, 18, 34, 35] (Figure 4; see Supplementary Information). The records are highly heterogeneous, with the platform and foreslope displaying distinctly different patterns. One view posits that cap carbonates initially formed with a uniform geochemistry and achieved heterogeneity through variable levels of aragonite neomorphism in the presence of meltwater and seawater, due to differences in cross-platform hydrological conditions. Our new results, however, indicate that the heterogeneity is a primary consequence of precipitation from these waters and their mixtures, which necessarily has different implications for interpreting geochemical signals within cap carbonates. Several studies have observed that epi-platform cap dolostones display synchronous negative $\delta^{44/40}$Ca excursions of varying magnitudes[16-18, 35]. We interpret this as evidence that global deglaciation drove local differences in seawater-meltwater mixing and carbonate precipitation rates.

Overall, the $\delta^{44/40}$Ca-$\delta^{88/86}$Sr multi-proxy provides a novel perspective on the origin of Marinoan cap carbonates. Detection of the "kinetic slope" in the Arbeitsgenot dolostone offers compelling evidence for primary signal preservation. The result implies that all $\delta^{44/40}$Ca and $\delta^{88/86}$Sr values, including the highest ones, are primary. We find that high $\delta^{44/40}$Ca values are due to relatively slow rates of primary carbonate formation rather than the secondary addition of Ca during fluid-buffered alteration. The highest $\delta^{44/40}$Ca and $\delta^{88/86}$Sr values for the Arbeitsgenot dolostone could provide evidence that dolomite precipitates more slowly than calcite. At a minimum, the data suggest that dolomitization during and following the glacial meltdown did not reset initial $\delta^{44/40}$Ca and $\delta^{88/86}$Sr values. Moreover, the Arbeitsgenot dolostone must have been deposited at a rate faster than the rate of seawater-meltwater mixing, which numerical simulations[47] and geochronological data[50] suggest was on the order of $10^5$ years or less. Overall, rapid formation of cap carbonate sequences may have sequestered excess atmospheric $CO_2$ that accumulated during the glaciation, leading to stabilization of Neoproterozoic climate[25]. A combination of seawater-meltwater mixing and kinetic isotope effects satisfactorily explains all $\delta^{44/40}$Ca and $\delta^{88/86}$Sr data presented herein, which indicates that any diagenetic overprinting was insufficient to obscure the primary signal. As these mechanisms also have potential to affect the carbon isotope composition of marine carbonates[10, 51], they could have contributed to the negative $\delta^{13}$C excursion following the Marinoan Snowball Earth event[4, 34].

Our findings offer a conceptual picture where global deglaciation amplified local coastal processes by intensifying riverine runoff and enhancing continental chemical weathering inputs, which shifted the isotopic composition of local seawater and accelerated carbonate precipitation rates. As these processes are expected to spatially and temporally vary for numerous reasons inherent to coastal settings, they provide a simple answer for the long-recognized geochemical heterogeneity of Neoproterozoic cap carbonates[9, 16, 18, 34, 45].

More broadly, our study illustrates the power of the $\delta^{44/40}$Ca-$\delta^{88/86}$Sr multi-proxy for probing the rock record. Detection of the "kinetic slope" in an inorganic Neoproterozoic dolostone (this study) and a biological Early Cretaceous limestone[22] provides a basis to argue that fundamentally different types of carbonates form in similar ways and are more robust archives of primary signals than widely perceived. As carbonate precipitation rates are proportional to $p$CO$_2$ via [CO$_3^{2-}$], these findings point the way for developing $\delta^{44/40}$Ca and $\delta^{88/86}$Sr as sensitive proxies for climate and saturation state.

**Methods**

We examined the same suite of samples from the Arbeitsgenot (n = 21) and Ombaatjie (n = 16) sections as reported by Hurtgen et al.[33] Sample preparation, elemental analyses, and Ca and Sr isotope measurements followed the same protocols described in detail by Wang et al.[49] Briefly, approximately 200 mg of powdered sample were dissolved in ~10 mL of ultrapure 5% HNO$_3$ in acid-cleaned centrifuge tubes, and the mixtures were gently shaken overnight to ensure complete reaction. The mixtures were centrifuged and passed through acid-cleaned 0.45 μm polypropylene syringe filters, collected in Teflon beakers, dried at 90°C, and then re-dissolved in 15 mL of 5% HNO$_3$ for elemental and isotopic analyses.

Sub-samples (Arbeitsgenot: n = 5; Ombaatjie: n = 5) were selected for a sequential leaching experiment, modified from previous studies[49], to test whether $^{87}$Sr/$^{86}$Sr ratios vary as a function of acid strength. First, a 1N ammonium acetate solution (NH$_4$Ac, buffered to pH=8.2) was applied to isolate the "exchangeable fraction," and the residues were then sequentially reacted with acetic acid (HAc) having concentrations of 0.25%, 1%, and 5%. For each step, the initial sample and remaining residues were reacted with ~25 mL of solution in acid-cleaned 50 mL centrifuge tubes, and the mixtures were agitated for 15 min and then centrifuged. The leachates were filtered through 0.45 μm polypropylene syringe filters, collected in Teflon beakers, dried at 90°C, and then re-dissolved in 15 mL of 5% $HNO_3$. The residues were rinsed with Milli-Q water between steps. Leachates for one sample from the Arbeitsgenot section (A-13), which yielded an unusually high bulk $^{87}Sr/^{86}Sr$ ratio and appears to be an outlier in Figure 3, were further analyzed for $\delta^{44/40}Ca$ and $\delta^{88/86}Sr$ values.

Elemental concentrations (Al, Ba, Ca, Fe, K, Mg, Mn, Na, Sr) were measured with a Thermo Scientific iCAP 6500 ICP-OES at Northwestern University. The samples were diluted to minimize matrix effects. The analytical standard NIST 1643f was repeatedly analyzed to assess the accuracy and reproducibility of the method. Results were within ±5% of reported concentrations.

Ca and Sr isotope ratios were measured in the Radiogenic Isotope Geochemistry Clean Laboratory at Northwestern University using a Thermo-Fisher Triton Multi-Collector Thermal Ionization Mass Spectrometer equipped with $10^{11}$ Ω amplifier resistors. Ca isotope ratios ($^{44}Ca/^{40}Ca$) were measured using an optimized $^{43}Ca$-$^{42}Ca$ double-spike technique. Aliquots of dissolved samples containing 50 µg of Ca were weighed into Teflon vials, spiked, capped, and heated at ~60°C overnight to ensure complete sample-spike equilibration. The solutions were dried, re-dissolved in 0.5 mL of 1.55N HCl, and eluted through Teflon columns packed with Bio-Rad AG MP-50 cation exchange resin to separate Ca from K and other matrix elements. The purified Ca fractions were dried, reacted with 35% $H_2O_2$ to oxidize potential organic compounds, and re-dried. Ca fractions were re-dissolved in concentrated $HNO_3$, dried, and finally, re-dissolved in 5 µL of 3N $HNO_3$. Approximately 12.5 µg of Ca, followed by 0.5 µL of 10% $H_3PO_4$, was loaded onto outgassed, single Ta filament assemblies. The $^{41}K$ beam (<0.0001V) was monitored to confirm that $^{40}K$ did not isobarically interfere with $^{40}Ca$ ($^{40}K/^{41}K$ = 0.00174). All results are reported in δ-notation relative to the OSIL Atlantic Seawater standard (SW), where $\delta^{44/40}Ca$ (‰)

= [($^{44}$Ca/$^{40}$Ca)$_{smp}$/($^{44}$Ca/$^{40}$Ca)$_{SW}$−1]×1000. The double-spike was frequently recalibrated by measuring at least 6 SW standards and 2 NIST 915b standards every 30 or fewer samples. During the period of study, repeated analyses yielded $\delta^{44/40}$Ca$_{SW}$ = 0.000±0.003‰ (2σ$_{SEM}$, n=26) and $\delta^{44/40}$Ca$_{915b}$ = -1.134±0.005‰ (2σ$_{SEM}$, n=8). The current long-term values for the laboratory are $\delta^{44/40}$Ca$_{SW}$ = 0.000±0.002‰ (2σ$_{SEM}$, n=635) and $\delta^{44/40}$Ca$_{915b}$ = -1.133±0.003‰ (2σ$_{SEM}$, n=257). These results correspond to a long-term, external reproducibility (2σ$_{SD}$) of ±0.05‰, which is the uncertainty assigned to samples.

The chemical weathering of old, K-rich crustal rocks could have supplied excess $^{40}$Ca produced by the radioactive decay of $^{40}$K, thereby lowering carbonate $\delta^{44/40}$Ca values. To test this hypothesis, we measured $\delta^{44/42}$Ca values for a subset of samples having low $\delta^{44/40}$Ca values and high $^{87}$Sr/$^{86}$Sr ratios (n = 9), using a $^{48}$Ca-$^{43}$Ca double-spike protocol[52]. $^{44}$Ca/$^{42}$Ca ratios are reported in δ-notation (in ‰) relative to OSIL SW, where $\delta^{44/42}$Ca (‰) = [($^{44}$Ca/$^{42}$Ca)$_{smp}$/($^{44}$Ca/$^{42}$Ca)$_{SW}$−1]×1000. During the period of study, repeated analyses of standards yielded $\delta^{44/42}$Ca$_{sw}$ = 0.000 ± 0.022‰ (2σ$_{SEM}$, n=8) and $\delta^{44/42}$Ca$_{915b}$ = -0.596 ± 0.023‰ (2σ$_{SEM}$, n = 4). The short-term reproducibility (2σ$_{SD}$) for the session is better than the long-term external reproducibility documented for the method (±0.126‰). The latter value is adopted as the uncertainty, and samples were run in triplicate and averaged, resulting in a standard error (2σ$_{SEM}$) of ±0.073‰.

For $^{87}$Sr/$^{86}$Sr measurements, aliquots of dissolved samples containing ~450 ng of Sr were weighed into Teflon vials, dried, re-dissolved in 8M HNO$_3$, and eluted through inverted pipet tips packed with Eichrom Sr-Spec resin. The purified Sr fractions were dried and re-dissolved in 3 μL of 3N HNO$_3$. Approximately ~150 ng of Sr was loaded on outgassed, single Re filament assemblies together with 1 μL of a TaCl$_5$ solution. $^{87}$Sr/$^{86}$Sr ratios were measured using multi-dynamic mode. The $^{85}$Rb beam was monitored to ensure that $^{87}$Rb did not isobarically interfere with $^{87}$Sr. Instrumental mass fractionation was corrected by normalizing $^{86}$Sr/$^{88}$Sr ratios to a fixed value of 0.1194, using an exponential law. During the period of study, repeated measurements of NBS 987 yielded a mean $^{87}$Sr/$^{86}$Sr ratio of 0.710250±0.000002 (2σ$_{SEM}$, n=12). The current, long-term mean $^{87}$Sr/$^{86}$Sr ratio for this method is 0.710251±0.000001 (2σ$_{SEM}$, n=231). These results correspond to a long-term external reproducibility (2σ$_{SD}$) of ±0.000010, which is the uncertainty assigned to samples.

$^{88}$Sr/$^{86}$Sr ratios were measured using an $^{87}$Sr-$^{84}$Sr double-spike technique. Samples containing ~600 ng Sr were weighted into Teflon vials, spiked, capped, and heated at ~60°C overnight. The solutions were dried, re-dissolved in 8N HNO$_3$, and purified using the same elution procedure as for $^{87}$Sr/$^{86}$Sr. After drying, the purified Sr fractions were re-dissolved in 4.5 µL of 3N HNO$_3$, and 1.5 µL aliquots containing ~200 ng of Sr were loaded on outgassed, single Re filament assemblies together with 1 µL of a TaCl$_5$ solution. During each measurement, the $^{85}$Rb beam was monitored to confirm that $^{87}$Rb did not isobarically interfere with $^{87}$Sr. The data reduction included input of sample $^{87}$Sr/$^{86}$Sr ratios. All $^{88}$Sr/$^{86}$Sr ratios are reported in δ-notation relative to NBS 987, where δ$^{88/86}$Sr (in ‰) = [($^{88}$Sr/$^{86}$Sr)$_{smp}$/($^{88}$Sr/$^{86}$Sr)$_{NBS987}$−1]×1000. The double-spike was frequently recalibrated by measuring at least 6 NBS 987 standards and 2 SW standards every 30 or fewer samples. During the period of study, repeated analyses yielded δ$^{88/86}$Sr$_{NBS987}$ = 0.000±0.002‰ (2σ$_{SEM}$, n=25) and δ$^{88/86}$Sr$_{SW}$ = 0.397±0.003‰ (±2σ$_{SEM}$, n=8). The current, long-term external reproducibility for NBS 987 is 0.000±0.001‰ (±2σ$_{SEM}$, n=297), and the current, long-term external reproducibility for seawater is 0.396±0.001‰ (±2σ$_{SEM}$, n=152). The uncertainty assigned to samples is ±0.020‰ (2σ$_{SD}$).

Strict protocols were adopted to confirm data quality and reproducibility. For Ca isotope measurements, total procedural blanks determined with a $^{43}$Ca isotope dilution method were 50–80 ng (n = 6), which is negligible compared to the amount of sample Ca processed. For Sr isotope measurements, total procedural blanks determined with an $^{84}$Sr isotope dilution method were ~100 pg (n = 5), which is also negligible compared to the amounts of sample Sr processed for both analyses. Samples were analyzed out of stratigraphic order spanning a period of three years. Approximately 40% (n = 14) of the samples were randomly chosen for duplicate measurements. Two approaches were taken to measure duplicates, including reanalysis of the same solution of dissolved powder and re-dissolution and reanalysis of the same powder. As shown in Supplementary Table 1, all duplicates agreed with original measurements, within the specified uncertainties.

## Acknowledgements

We thank M. E. Ankney and A. Masterson for laboratory assistance. This work was supported by a David and Lucile Packard Foundation Fellowship (2007-31757) and a National Science Foundation grant (NSF-EAR 0723151) awarded to A.D.J., and a National Aeronautics and Space Administration grant (NASA- 80NSSC17K0245) awarded to M.T.H., A.D.J., and B.B.S.

## Data availability

All data analyzed in this study are included in the Supplementary Information files. Source data are provided with this paper.

## Author contributions

48. K. J. Huang, *et al.*, Episode of intense chemical weathering during the termination of the 635 Ma Marinoan glaciation. *Proc. Natl. Acad. Sci. U. S. A.* **113**, 14904–14909 (2016).

49. J. Wang, *et al.*, Coupled $\delta^{44/40}$Ca, $\delta^{88/86}$Sr, and $^{87}$Sr/$^{86}$Sr geochemistry across the end-Permian mass extinction event. *Geochim. Cosmochim. Acta* **262**, 143–165 (2019).

50. C. Zhou, M. H. Huyskens, X. Lang, S. Xiao, Q. Z. Yin, Calibrating the terminations of Cryogenian global glaciations. *Geology* **47**, 251–254 (2019).

51. J. V. Turner, Kinetic fractionation of carbon-13 during calcium carbonate precipitation. *Geochim. Cosmochim. Acta* **46**, 1183–1191 (1982).

52. G. O. Lehn, A. D. Jacobson, Optimization of a $^{48}$Ca-$^{43}$Ca double-spike MC-TIMS method for measuring Ca isotope ratios ($\delta^{44/40}$Ca and $\delta^{44/42}$Ca): limitations from filament reservoir mixing. *J. Anal. At. Spectrom.* **30**, 1571–1581 (2015).
21

**Figures and figure captions**

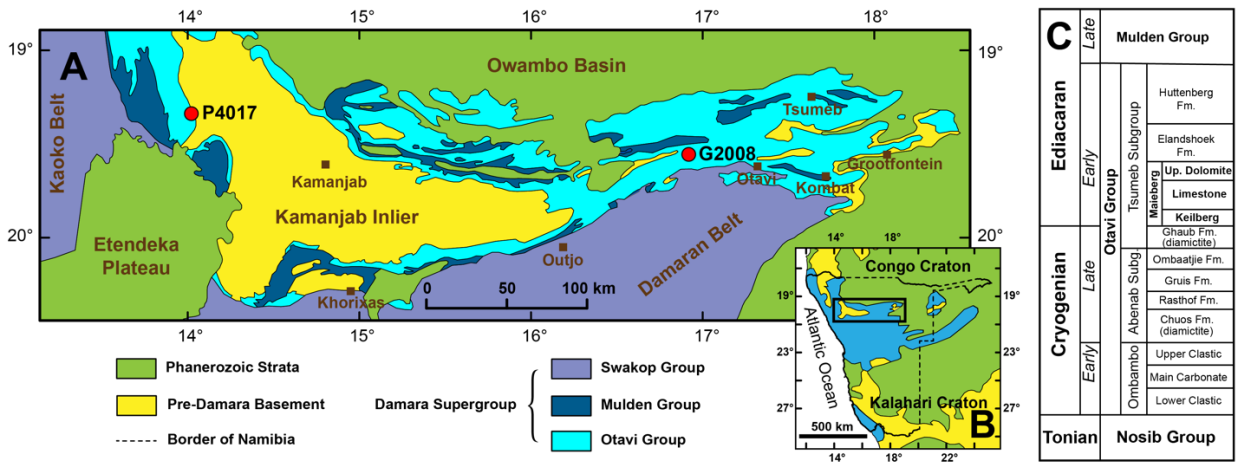

**Figure 1** Geologic map of Namibia (A and B), locations of studied sections[6] (shown in red circles), and stratigraphic column of Otavi Group (C).

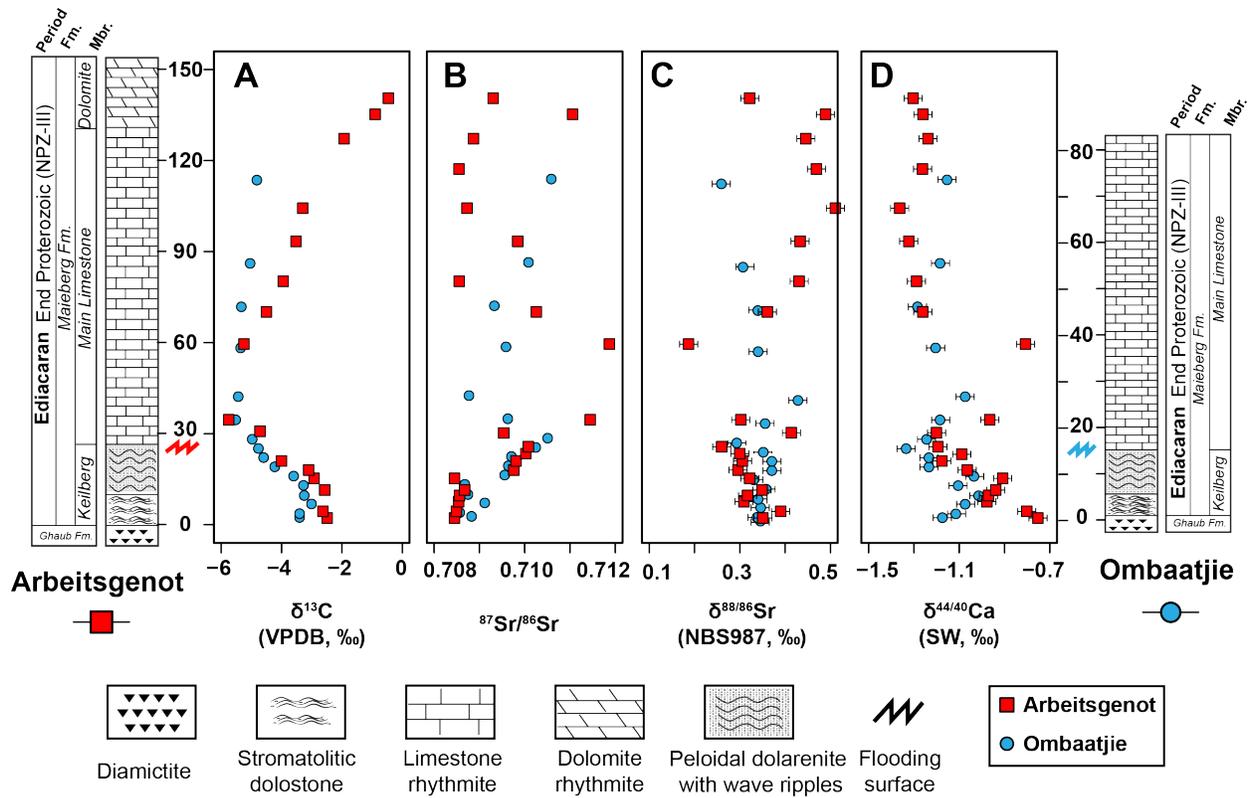

**Figure 2** Stratigraphic correlation of (A) $\delta^{13}C$, (B) $^{87}Sr/^{86}Sr$, (C) $\delta^{88/86}Sr$, and (D) $\delta^{44/40}Ca$ records of the Maieberg Formation (cap carbonate sequence) from the Arbeitsgenot and Ombaatjie sections, Namibia (the data series are superimposed).

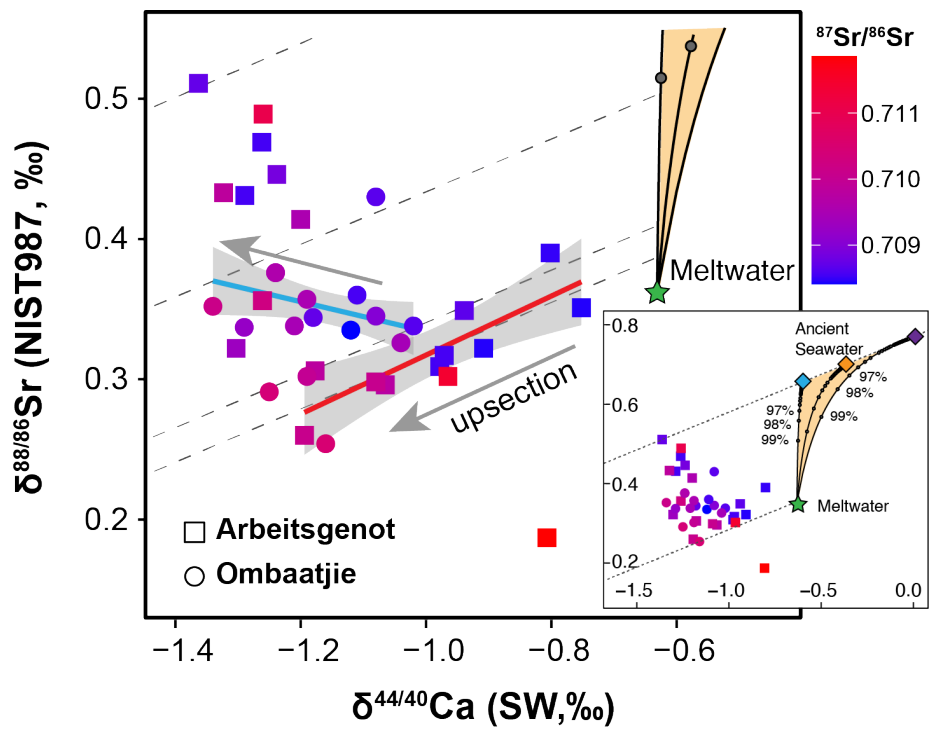

**Figure 3** Cross-plot of $\delta^{44/40}Ca$ and $\delta^{88/86}Sr$ values color-contoured with $^{87}Sr/^{86}Sr$ ratios. Arrows indicate the direction of stratigraphically upward in the cap dolostones. The grey shaded regions represent the 95% confidence intervals for linear regressions performed only on cap dolostone samples from the Arbeitsgenot section (solid red line, $R^2 = 0.74$, $p$-value $= 1.5 \times 10^{-3}$) and the Ombaatjie section (solid blue line, $R^2 = 0.41$, $p$-value $= 6.4 \times 10^{-2}$). All other datapoints reflect cap limestone samples from the two sections. Solid black lines show theoretical mixtures between meltwater (green star) and various seawater end-members, which may have been brackish or even a brine. The blue, orange, and purple diamonds reflect different scenarios for the ancient seawater end-member. Yellow area denotes the potential range of mixture compositions, and green circles show proportions of meltwater in seawater-meltwater mixtures, at 1% increments. See Supplementary Information for more details. Grey dashed lines are mass-dependent fractionation lines with slopes of 0.19, which is the theoretical value expected for kinetic isotope fractionation of $Ca^{2+}$ and $Sr^{2+}$ six-fold coordinated aquocomplexes during dehydration at the carbonate-water

interface[27]. Lower carbonate $\delta^{44/40}$Ca and $\delta^{88/86}$Sr values correspond to faster precipitation rates, while higher values correspond to slower precipitation rates. Other studies of synthetic and natural samples have detected a "kinetic slope" of 0.19[22, 27]. Kinetic fractionation during cation diffusion through a boundary layer surrounding growing crystals should produce a slope of 0.24[27]. Of the four existing studies to report paired $\delta^{44/40}$Ca-$\delta^{88/86}$Sr measurements, none have provided clear evidence for diffusional control[22, 27, 41, 49], although contributions from this mechanism would not change the interpretation offered here. Finally, no evidence is observed for equilibrium control, which confirms previous assertions that Ca and Sr isotope fractionation during most carbonate formation is kinetic[27].

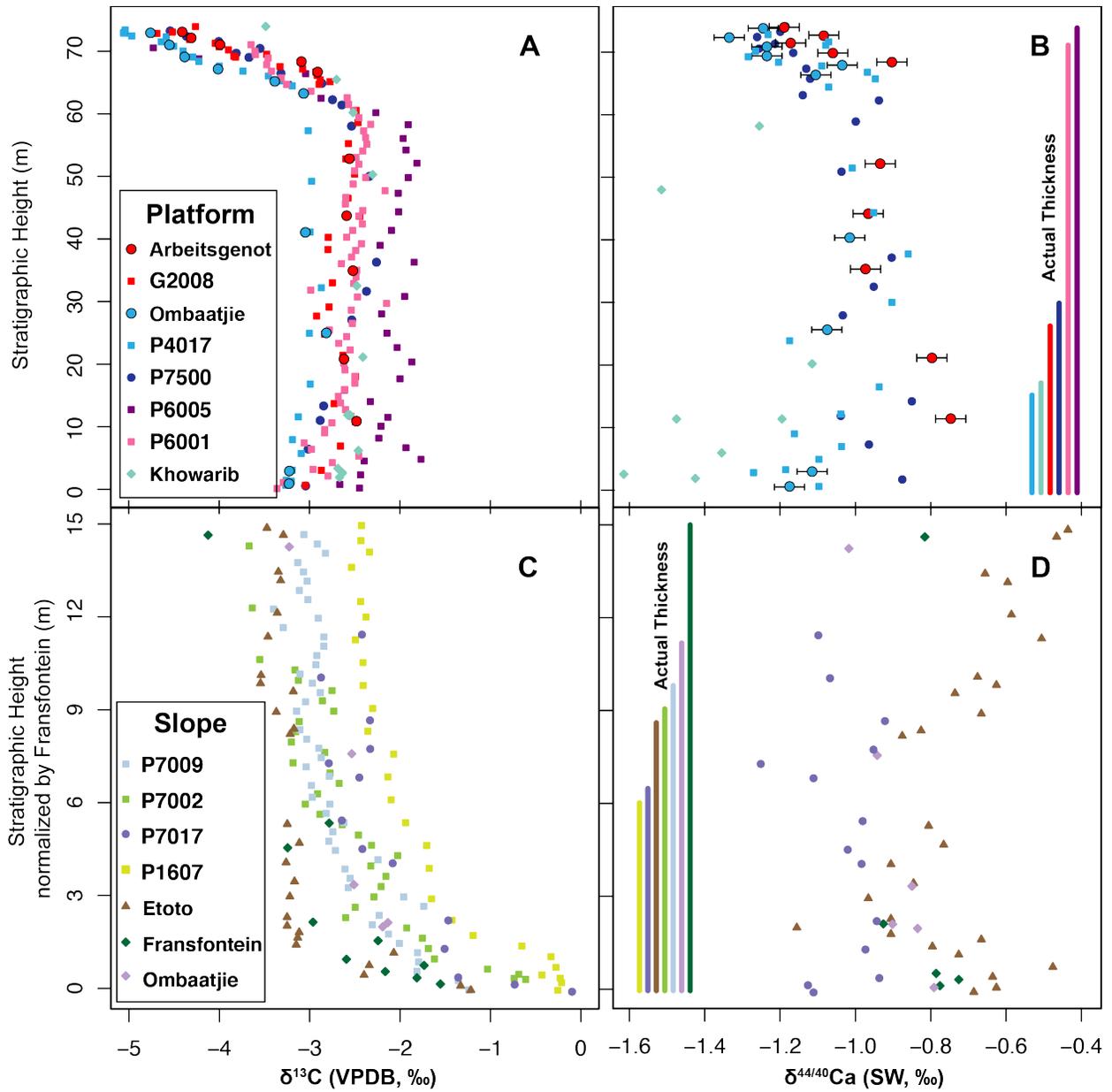

**Figure 4** Compilation of $\delta^{13}C$ and $\delta^{44/40}Ca$ records[16, 18, 35] from sections across the Namibian Otavi platform (A and B; normalized to P6005) and its foreslope (C and D; normalized to Fransfontein). Stratigraphic correlations utilized $\delta^{13}C$ chemostratigraphy following Hoffman et al.[34] See Supplementary information for more details.